\documentclass{article}

\usepackage{PRIMEarxiv}

\usepackage[utf8]{inputenc} 
\usepackage[T1]{fontenc}    
\usepackage{hyperref}       
\usepackage{url}            
\usepackage{booktabs}       
\usepackage{amsfonts}       
\usepackage{nicefrac}       
\usepackage{microtype}      
\usepackage{lipsum}
\usepackage{fancyhdr}       
\usepackage{graphicx}       
\graphicspath{{media/}}     
\usepackage{multirow}
\usepackage{subcaption}
\usepackage{siunitx}
\usepackage[most]{tcolorbox}
\usepackage{xcolor} 
\usepackage{algorithm}
\usepackage{algpseudocode}
\usepackage{natbib}
\tcbset{
  promptbox/.style={
    enhanced,
    colback=gray!10,        
    colframe=gray!80,       
    coltitle=white,         
    colbacktitle=gray!50,   
    fonttitle=\bfseries\footnotesize,
    fontupper=\footnotesize,
    boxrule=0.5pt,          
    titlerule=0.5pt,        
    titlerule style={gray!80}, 
    arc=3pt,                
    outer arc=3pt,
    top=4pt, bottom=4pt,
    left=6pt, right=6pt,
    toptitle=2pt, bottomtitle=2pt, 
    before=\par\nobreak,    
    width=\columnwidth       
  }
}

\title{A Hierarchical Agentic Framework for Autonomous Drone-Based Visual Inspection
}

\author{
  Ethan Herror, Xian Yeow Lee, Gregory Sin, Teresa Gonzalez Diaz, Ahmed Farahat, Chetan Gupta \\
  Hitachi America Ltd., R\&D, \\
  Santa Clara \\
  \{ethan.herron, xian.lee, gregory.sin, teresa.gonzalezdiaz, ahmed.farahat, chetan.gupta\}@hal.hitachi.com
  }

\begin{document}
\maketitle

\begin{abstract}
Autonomous inspection systems are essential for ensuring the performance and longevity of industrial assets. 
Recently, agentic frameworks have demonstrated significant potential for automating inspection workflows but have been limited to digital tasks. 
Their application to physical assets in real-world environments, however, remains underexplored. 
In this work, our contributions are two-fold: first, we propose a hierarchical agentic framework for autonomous drone control, and second, a reasoning methodology for individual function executions which we refer to as ReActEval. 
Our framework focuses on visual inspection tasks in indoor industrial settings, such as interpreting industrial readouts or inspecting equipment. 
It employs a multi-agent system comprising a head agent and multiple worker agents, each controlling a single drone. 
The head agent performs high-level planning and evaluates outcomes, while the worker agents implement our ReActEval methodology to reason over and execute low-level actions. 
Operating entirely in the natural language space, ReActEval follows a plan, reason, act, evaluate cycle, enabling drones to handle tasks ranging from simple navigation (e.g., flying forward 10 meters and land) to complex high-level tasks (e.g., locating and reading a pressure gauge).
The evaluation phase serves as a feedback and/or replanning stage, ensuring the actions executed align with user objectives while preventing undesirable outcomes. 
We evaluate the framework in a simulated environment with two worker agents, assessing performance qualitatively and quantitatively based on task completion across varying levels of task complexity and agentic workflow efficiency. 
By leveraging natural language processing for agent communication, our approach offers a novel, flexible, and user-accessible alternative to traditional drone-based solutions, enabling a more autonomous problem-solving approach to industrial inspection tasks without requiring extensive user intervention.
\end{abstract}


\section{Introduction}

Industrial inspection systems face rising demand for autonomous solutions capable of reducing safety risks and operational costs while maintaining coverage of critical infrastructure. Manual inspection in hazardous environments, such as chemical plants and power facilities, introduces safety concerns and limits the frequency and thoroughness of assessments due to human constraints(~\cite{osha}).

Current drone-based inspection systems rely heavily on skilled operators and manual intervention(~\cite{RODRIGUEZ2024102330}). They use pre-programmed flight paths that lack adaptability for dynamic industrial environments. Real-time pilot requirements create decision-making bottlenecks, while multi-drone coordination presents significant cognitive load challenges. These limitations prevent current systems from scaling effectively across three critical dimensions: task breadth (deployment in diverse industrial settings), task complexity, and the number of concurrently deployed drones.

Agentic frameworks, systems composed of multiple coordinated Large Language Models (LLMs) that operate together to achieve predefined tasks, have shown remarkable success in digital domains, particularly in software development with agentic Integrated Development Environments (IDEs) and coding agents(~\cite{ai_coding, wu2023autogen}) as well as in supporting researchers in performing scientific research(~\cite{gridach2025agentic, huang2025deep}). These systems use natural language interfaces and multi-agent coordination to achieve human-level reasoning in complex problem-solving scenarios, suggesting significant potential for physical applications.

However, applying multi-agent systems to physical asset inspection reveals critical research gaps. This work addresses two fundamental challenges. First, how should multiple agents, in our case drones, be managed within the system? Second, how should each drone's agent handle task execution effectively?

To address multi-agent management, we introduce a hierarchical agentic framework in which a single head agent generates high-level plans for individual worker agents, which are LLMs controlling their respective drones. This architecture centralizes communication between frontline workers such as technicians, operators, inspectors and the broader inspection system, reducing latency and eliminating the need for complex inter-agent coordination.

Initial experimentation with the hierarchical framework revealed poor performance in drone-level task execution, including incorrect movements and incomplete task completion. These findings motivated the development of our second contribution: the \textbf{ReActEval} framework. Although the effectiveness of various reasoning methods for LLM-driven vehicle control remains poorly understood, the foundational ReAct framework by~\citet{yao2022react} provides an established approach that combines reasoning and acting through interleaved generation of reasoning traces and task-specific actions. Building on this established framework, we propose \textbf{ReActEval}, which extends ReAct by adding a critical third step—evaluate—after each action execution to enable structured self-correction in physical tasks, addressing the unique challenges of real-world drone control.

We evaluate our ReActEval method within the hierarchical multi-agent framework for autonomous drone visual inspection. Through systematic comparison against ReAct and a simplified Act method across different model capabilities and task complexities, we provide the first comprehensive analysis of how reasoning method selection affects performance in physical agentic systems. Our contributions are: (1) a hierarchical agentic framework for drone-based industrial vision tasks, such as inspection, monitoring, tracking and identification, (2) the novel ReActEval framework for enhanced planning and evaluation in physical agentic systems, and (3) a systematic comparison of reasoning approaches across different model capabilities that reveals important performance tradeoffs.

Our findings reveal counterintuitive relationships between reasoning methods and model capability. We demonstrate that more complex reasoning methods are not universally superior, but rather their effectiveness depends critically on the underlying model's capability and task complexity. This challenges the assumption that sophisticated reasoning frameworks always improve performance. These insights offer guidance in developing effective autonomous inspection systems and lay the groundwork for future research on agentic frameworks for physical tasks.

\section{Related Work}

Recent advancements in agentic frameworks have shown potential for complex problem-solving in digital domains (~\cite{elrefaie2025aiagentsengineeringdesign, zou2025elagenteautonomousagent, pandey2025openfoamgptragaugmentedllmagent}). The application of these frameworks, particularly those driven by Large Language Models (LLMs), to robotics and physical systems is an emerging area of research(~\cite{chen2025multiagentsystemsroboticautonomy, zhang2024bridgingintelligenceinstinctnew}). The intersection of Unmanned Aerial Vehicles (UAVs) and LLMs has emerged as a particularly promising domain, with multiple recent works exploring various aspects of autonomous aerial intelligence.

\subsection{LLM-Driven UAV Control Frameworks}

Several recent works have proposed frameworks for integrating LLMs with UAV control systems. ~\citet{Tian2025} provide a comprehensive overview of LLM-UAV integration, introducing the concept of "agentic low-altitude mobility" where UAVs make intelligent decisions in complex environments through multi-modal capabilities combining vision, language, and sensor fusion. Their analysis identifies key challenges including computational constraints, real-time processing requirements, and safety-critical decision making.

\citet{wang2025gscepromptframeworkenhanced} addressed the critical safety and reliability concerns in LLM-UAV integration by proposing the GSCE (Goal, State, Context, Execute) prompt framework. Their structured approach to prompt engineering significantly improves reliability of LLM-driven drone control and demonstrates the importance of systematic reasoning processes for safety-critical UAV operations.

\subsection{Multi-Domain Applications and Reasoning Paradigms}

\citet{sapkota2025uavsmeetagenticai} provide a comprehensive multidomain survey of agentic AI applications in UAV systems, establishing a taxonomy of autonomous aerial intelligence capabilities across diverse domains including agriculture, surveillance, and search and rescue. Their analysis reveals that different domains require specialized adaptations of core agentic capabilities, with current limitations including safety assurance, regulatory compliance, and ethical considerations.

\citet{javaid2024largelanguagemodelsuavs} assess the current state of LLM applications in UAV systems and identify key technological pathways for future development. Their technology readiness assessment reveals that while LLM technology shows significant promise for UAV applications, implementation challenges persist around model optimization, edge processing, deployment, and integration with existing UAV systems.

\subsection{Research Gaps and Reasoning Frameworks}

Despite these advances, several critical gaps remain in the literature. Current works primarily focus on system architecture and domain-specific applications, but lack systematic evaluation of different reasoning methods for LLM-driven robot control in physical environments. The effectiveness of various reasoning approaches, from simple direct action to complex multi-step reasoning, remains poorly understood, particularly regarding the tradeoffs between performance, computational cost, and task complexity.

Given the emerging nature of this field, few established reasoning frameworks exist for physical agentic systems. The foundational work of \citet{yao2022react} introduced the ReAct framework, which synergizes reasoning and acting in language models through interleaved generation of reasoning traces and task-specific actions. Their approach demonstrated that reasoning traces help models induce, track, and update action plans while handling exceptions, and actions allow interfacing with external sources to gather additional information. ReAct showed significant improvements on question answering, fact verification, and interactive decision-making tasks, establishing the value of combining reasoning and acting in a unified framework. While alternative approaches exist, such as the GSCE prompt framework(~\cite{wang2025gscepromptframeworkenhanced}) which focuses on structured prompt engineering rather than reasoning-action cycles, ReAct represents the most relevant framework for our domain given the limited alternatives in physical systems. Moreover, agentic systems are inherently flexible, with methods typically tailored to specific tasks rather than following standardized approaches. This flexibility necessitates systematic evaluation across different conditions to understand when and why particular methods succeed.

Prior work has explored LLM-based planning for robotic tasks, demonstrating the ability to translate high-level natural language instructions into executable actions(~\cite{wang2024large}). However, the question of how different reasoning structures affect performance across varying model capabilities and task complexities has not been systematically addressed. This gap is particularly important given the computational constraints and safety requirements inherent in UAV operations.

Our work builds on these foundations by introducing the hierarchical agentic framework and evaluating the ReActEval framework in a real-world drone inspection scenario. This approach addresses the identified gap in understanding how reasoning method selection interacts with model capability to affect overall system performance in physical domains.

\begin{figure}
    \centering
    \includegraphics[width=0.7\linewidth]{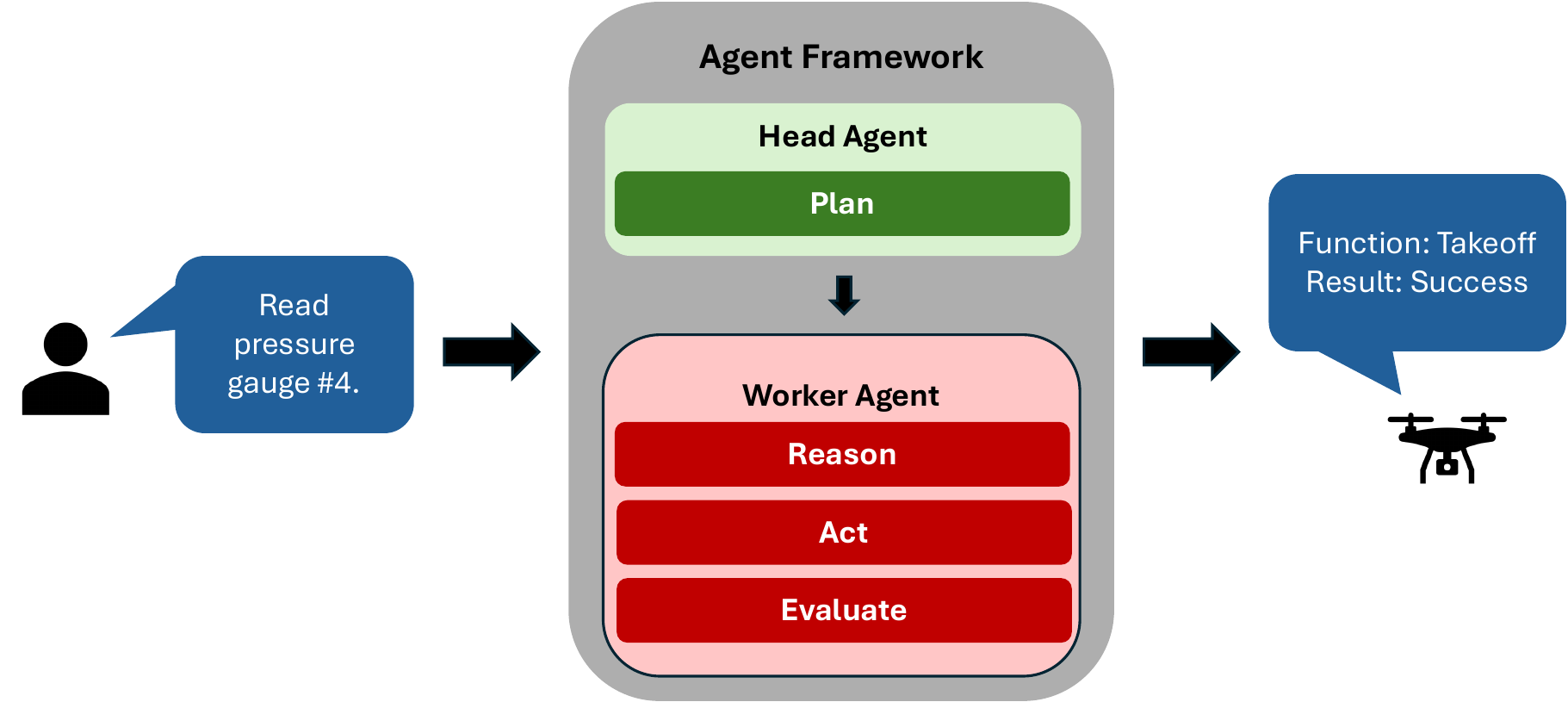}
    \caption{An overview of the hierarchical agentic framework. Users define a task for the agentic framework to complete with the drone. The Head Agent creates a plan to accomplish the user-defined task. The Worker Agent uses the \emph{ReActEval} method to execute actions that accomplish the high level plan defined by the Head Agent. The Worker Agent communicates directly with the drone and other tools (VLMs, secondary task-specific models, etc.) by directly calling API functions.}
    \label{fig:framework}
\end{figure}

\section{Methodology}
\label{sec:methods}

\subsection{Hierarchical Agentic Framework Architecture}

Our framework, shown in Fig.~\ref{fig:framework}, uses a hierarchical, multi-agent architecture consisting of a head agent for high-level planning and multiple worker agents for task execution. This structure employs one head agent controlling multiple worker agents, with each worker agent dedicated to a single drone, offering several appealing and practical benefits.

First, this architecture enables the framework to seamlessly scale across an arbitrary numbers of drones without requiring structural modifications to the control system. The number of available drones is specified by the user during system initialization, and the head agent dynamically allocates tasks across these predefined drones based on the current request.

Second, the user-to-head-agent-to-worker-agent communication pathway enables consistent task description to worker agents, leading to more reliable performance. Since users naturally prompt and describe tasks in different ways, the head agent serves as a standardization layer, translating diverse user inputs into consistent, structured task descriptions that worker agents can reliably interpret and execute.

Third, this hierarchical structure helps avoid context bloat through differentiated memory management. The framework operates over the course of a \emph{session}, defined as a series of multiple different tasks or user interactions. During a session, the head agent maintains a comprehensive running log of all tasks completed across the entire conversation history. In contrast, each worker agent's history resets after completing its assigned task (referred to as a \emph{thread}), preventing the accumulation of irrelevant historical context that could degrade performance on subsequent tasks.

More specifically, we distinguish between \emph{session history} and \emph{thread history}. Session history encompasses the complete record of user interactions, head agent planning decisions, and high-level task outcomes maintained by the head agent throughout the entire operational session. Thread history refers to the task-specific execution record maintained by individual worker agents, including their reasoning steps, actions taken, and evaluations performed during the completion of a single assigned sub-task.

The Head Agent processes user input through its \textit{Plan} function, which outputs a structured dictionary containing task allocations for specific drones. Each drone entry in this dictionary includes: (1) a step-by-step plan tailored to that drone's capabilities and current state, (2) an expected outcome describing the desired end state upon task completion, (3) an end flag boolean indicating whether or not to invoke a specific drone's worker agent, and (4) a response to the user providing feedback or status information. This structured output ensures consistent task specification and enables effective coordination across multiple drones. We formally define this process in Algorithm \ref{alg:head_agent}. Additionally, the prompt for the head agent's \textit{Plan} function is shown in the Appendix.

Worker Agents are responsible for controlling their respective drone through an iterative execution loop. This loop, with our proposed ReActEval approach, follows a structured process where agents first reason about the next required action, execute the corresponding function call, and then evaluate the result before proceeding. This iterative approach provides agents with enhanced flexibility and autonomy when solving assigned tasks, including the ability to self-terminate the execution loop during the evaluation phase upon task completion. Function calls are implemented through API tool calls, with drone-specific tools including \textit{Takeoff, Land, Move, Rotate} and \textit{Capture Image}. The tool-calling framework is agnostic to specific implementations, enabling seamless integration with arbitrary tools, e.g., vision-language models (VLM), custom analysis tools, or specialized functions required for given environment/task-set.

\begin{algorithm}
\caption{Multi Agent Drone Control}
\label{alg:head_agent}
\begin{algorithmic}[1]
\Procedure{ExecuteHeadAgent}{$user\_input$}
    \State $drone\_tasks \gets$ $\textsc{PLAN}(user\_input)$
    \State $results \gets \{\}$
    \For{each $agent\_id, task$ in $drone\_tasks$}
        \State $results[agent\_id] \gets \textsc{ReActEval}(task)$
    \EndFor
    \State $response \gets$ \parbox[t]{.75\linewidth}{$\textsc{Respond}(user\_input,$ \\ \hspace*{1.5em}$drone\_tasks, results)$}
    
    \State \Return $response$
\EndProcedure
\end{algorithmic}
\end{algorithm}

\subsubsection{ReActEval}
\label{subsection:ReActEval}
Our proposed ReActEval method, outlined in Algorithm \ref{alg:worker_agent}, uses a three-step "Reason-Act-Evaluate" process that provides worker agents with structured decision-making capabilities for autonomous drone control.

The Reason step serves as a planning phase for determining the next optimal function or action to execute. In this step, the agent reasons over the drone's current state (including position coordinates, heading, and other relevant parameters), the plan for the current sub-task provided by the head agent, the intended outcome for the current task, and the thread history. The thread history, when applicable, contains previous actions executed, evaluation outputs, and next step notes from prior evaluation phases. The Reason step outputs a structured dictionary with keys 'reason' and 'intended action', where 'reason' contains the logical justification for the chosen action and 'intended action' specifies the next recommended action to execute.

The Act step takes the drone's current state, the intended action from the reasoning phase, and the thread history as inputs. This step translates the high-level intended action into executable function calls, directly interfacing with the drone's control API or the specified tool's API to perform the specified operation.

The Evaluate step provides critical assessment and loop control functionality. The agent takes the overall plan, expected outcome for the current task, the recently executed action, and thread history as inputs. The Evaluate step outputs a structured dictionary containing: (1) 'evaluation' - reasoning over the last action executed, current drone state, and other relevant information to assess task progress and completion status; (2) 'end flag' - a boolean value that determines whether to terminate the execution loop (true if the task has been completed, false to continue the ReActEval loop); and (3) 'next steps notes' - guidance and suggestions for the subsequent reasoning step, recommending the next optimal action based on current progress and remaining objectives.

The prompts for the Reason, Act, and Evaluate steps can be found in the Appendix \ref{sec:appendix}.

\begin{algorithm}
\caption{ReActEval}
\label{alg:worker_agent}
\begin{algorithmic}[1]

\Procedure{ReActEval}{$task$}
    \State $history \gets \textsc{Initialize}(task)$
    \State $iteration \gets 0$
    \While{$\neg task.complete \land iteration < max\_iters$}
        \State $iteration \gets iteration + 1$
        \State $reasoning \gets \textsc{Reason}(task, history)$
        \State $action \gets \textsc{Act}(reasoning)$
        \State $evaluation \gets \textsc{Evaluate}(task, history, action)$
        \State $task.complete \gets evaluation.end\_flag$
        \State $history \gets$ \parbox[t]{.75\linewidth}{$\textsc{UpdateHistory}(reasoning,$ \\ $\;\;\;\;\;action, evaluation)$}

    \EndWhile
    \State \Return $history$
\EndProcedure

\end{algorithmic}
\end{algorithm}

\subsubsection{ReAct}
\label{subsection:ReAct}
For comparison, we use the ReAct method, outlined in Algorithm \ref{alg:worker_agent_react}, which follows a "Reason-Act" cycle. The Reason and Act steps are similar to ReActEval, except that we add the 'end flag' key from ReActEval's evaluate step to determine loop termination. This end flag is updated after the Reason step, unlike ReActEval where it updates after Evaluation, avoiding unnecessary function executions. We edited the Reason's prompt to include examples on how to use the 'end flag' key in its output for this added functionality; otherwise the Reason and Act prompts are the same as those used in the ReActEval method to maintain consistency during experimentation. 

\subsubsection{Act}
\label{subsection:Act}
As a second baseline, the Act method, outlined in Algorithm \ref{alg:worker_agent_act}, removes both Reason and Evaluation steps. To maintain a fair comparison of this method against the others, we made two changes to the Act step. First, since the Act step makes direct function calls, we created a new 'termination' function, which when called, breaks the current execution loop. This has the same functionality as setting the end-flag key in other methods to \emph{True}. Second, we extended the prompt to include the plan and expected outcome from the head agent. To maintain consistency across the other methods, we used the same relevant portions from the other method's Reason prompts.

\section{Experiments}

Our experiments evaluate the performance of the proposed ReActEval method against two baselines (ReAct and Act, as described in the Methodology section \ref{sec:methods}) across different levels of task complexity. To assess the impact of LLM model size and reasoning capabilities, we test four different LLMs: \texttt{GPT-4.1 Nano}, \texttt{GPT-4.1}, \texttt{o4-mini}, and \texttt{o3}. These models span a spectrum of reasoning abilities and computational efficiency: \texttt{GPT-4.1} is a state-of-the-art large language model known for advanced reasoning and understanding, while \texttt{GPT-4.1 Nano} is a lightweight variant optimized for faster responses with lower computational cost but somewhat reduced capacity. The \texttt{o4-mini} and \texttt{o3} models are smaller, more efficient architectures that balance performance and resource use. This diverse selection allows us to identify how each method’s performance scales with model intelligence and pinpoint any capability thresholds required for success.

We categorize tasks into three levels of complexity. \textbf{Easy} tasks involve basic single or two-step commands, per drone, such as takeoff, land, or movement to specified coordinates, with clearly defined objectives requiring minimal planning or coordination. \textbf{Medium} tasks consist of multi-step sequences with explicit actions specified in the prompt, assessing each method's ability to execute longer sequences of coordinated commands across multiple drones. \textbf{Hard} tasks involve complex, realistic inspection scenarios requiring precise navigation, multi-perspective scene analysis, and visual interpretation, such as locating and reading pressure gauges in industrial environments. The specific task descriptions used can be found in Table \ref{tab:prompts}.

\begin{table}[h]

\centering
\caption{Tasks by Complexity Level}
\label{tab:prompts}
\begin{tabular}{|p{1.6cm}|p{12.cm}|}
\hline
\textbf{Complexity} & \textbf{Tasks} \rule{0pt}{12pt} \\
\hline
\multirow{12}{*}{Easy} & 
\begin{minipage}[t]{12.cm}
\vspace{0.05cm}
\begin{itemize}
\item What are your responsibilities?
\item Can you take off both drones?
\item Can you move drone 1 forward 2m?
\item Can you take a picture with drone 2?
\item Can you analyze the image with drone 2?
\item Takeoff and then land both drones safely.
\item What is the current state of both drones?
\item Rotate both drones 180 degrees.
\vspace{0.05cm}
\end{itemize}
\end{minipage} \\
\hline
\multirow{9}{*}{Medium} & 
\begin{minipage}[t]{12.cm}
\vspace{0.05cm}
\begin{itemize}
\item Fly both drones in a trajectory of a square with length 3m.
\item Drone 2, can you turn around and take a picture, and then land safely?
\item Drone 1, fly forward 4m, take a picture and describe what you see.
\item Drone 1, move right 5m and then up 5m, turn around and land. Drone 2, move left 5m and then up 2m, rotate $90^{\circ}$ and then land.
\item Fly both drones in the trajectory of a triangle with length 5m, drone 1 should move left first and drone 2 should move right first.
\vspace{0.05cm}
\end{itemize}
\end{minipage} \\
\hline
\multirow{9}{*}{Hard} & 
\begin{minipage}[t]{12.cm}
\vspace{0.05cm}
\begin{itemize}
\item The drones are located in the middle of a 10m x 2m room. Use both drones to capture images of each corner of the room. Each drone should assess 2 corners.
\item Drone 2, navigate to the pressure gauge located at (4m, 18m, 6m) and return its status.
\item There is an object located at (3, 4, 5). Drone 1, describe the object from the left side. Drone 2, describe the object from the right side.
\vspace{0.05cm}
\end{itemize}
\end{minipage} \\
\hline
\end{tabular}
\end{table}

\subsection{System and Environment}
All experiments are conducted within a simulated environment that maintains each drone's state through a command-execution-based state management system. We track the 3D Cartesian coordinates, heading, gimbal angle (camera orientation), and last executed command for each drone. This set of values constitutes the \emph{Drone State}, which is updated after every function execution based on the specific command type and its parameters.

The state update rules are defined as follows: \textit{takeoff} commands set the drone's flight status to True and altitude to 1.0m; \textit{land} commands set flight status to False and altitude to 0.0m; \textit{move} commands calculate new coordinates using trigonometric functions based on the drone's current heading and the specified direction (forward, backward, left, right, up, down) and distance; \textit{rotate} commands update the heading by adding the rotation angle to the current heading; and \textit{move\_gimbal} commands directly update the gimbal angle within the constrained range of 0-90 degrees.

The framework supports both simulated and real drone operations. For simulated drones, state changes are calculated based on the command parameters and current state. For real drones, the system polls the actual drone's API to retrieve current flight status, altitude, heading, and gimbal angle after command execution, ensuring state consistency with the physical drone. 

We utilize two drones in our experiments to simplify experimentation while demonstrating multi-drone coordination capabilities, although our framework is designed to scale to arbitrary numbers of drones. The first drone initializes at the origin (0,0,0), and the second drone starts at (0,2,0), positioned 2 meters apart in the y-direction.

The Worker Agents are equipped with additional tools including VLMs and YOLO(~\cite{redmon2016you}) models for visual analysis and object detection during inspection tasks. These tools enable the drones to perform complex visual interpretation required for industrial inspection scenarios.

\subsection{Evaluation Metrics}
We evaluate performance using two primary metrics: task completion rate and execution time.

For task completion, we use different scoring approaches based on complexity level due to the nature of the tasks. For easy and medium tasks, we manually analyze each prompt to identify the required sequence of function calls, as these tasks have deterministic action sequences with clearly specified objectives. The 8 easy tasks require 14 total actions, while the 5 medium tasks require 36 total actions. We score each task by counting successful function call executions, awarding one point per correctly executed function. For example, the task "Can you take off both drones?" requires two takeoff function calls and scores a maximum of two points—one point for each successful takeoff execution.

Hard tasks involve complex, open-ended scenarios that can be accomplished through multiple valid approaches, making function-level scoring impractical. Instead, we manually decompose these tasks into higher-level subtasks and award points based on subtask completion. For instance, "Use both drones to capture images of each corner of the room. Each drone should assess 2 corners" scores four points total, one for each corner successfully imaged, regardless of the specific function sequence used to achieve the imaging. Note that in this type of the tasks, the instructions given by the user are also not necessarily explicit function calls and requires the LLM to decompose, delegate and/or convert a possibly ambiguous user instruction into a series of executable drone operations. 

In all cases, scoring follows a strict sequential evaluation where points are awarded only for correct function calls executed in the proper context. If an incorrect function is called, scoring stops at that point, and subsequent actions receive no credit even if executed correctly. In our simulated environment, correctly specified function calls always succeed, ensuring that failures result from incorrect function selection rather than execution errors.

We measure execution time as the end-to-end latency from when a user request is received until the final response is generated. This captures the complete pipeline including head agent planning, worker agent execution cycles using the respective method (ReActEval, ReAct, or Act), and final response generation across all assigned drones.

\section{Results}

\begin{table}[t]
\centering
\setlength{\tabcolsep}{4pt}
\caption{Performance Comparison Across Models and Difficulty Levels. Overall column shows accuracy across all complexity levels.}
\label{tab:task_results}
\begin{tabular}{llcccc}
\toprule
\textbf{Method} & \textbf{Model} & \textbf{Easy} & \textbf{Medium} & \textbf{Hard} & \textbf{Overall} \\
\midrule
\multirow{4}{*}{ReActEval} 
& GPT 4.1-nano & \textbf{14/14} & 13/36          &          2/13  &         0.460  \\
& GPT 4.1      &          13/14 & \textbf{34/36} &          4/13  &         0.810  \\
& o4-mini      & \textbf{14/14} & \textbf{34/36} &          6/13  &         0.857  \\
& o3           &          13/14 & \textbf{34/36} & \textbf{10/13} & \textbf{0.905} \\
\midrule
\multirow{4}{*}{ReAct} 
& GPT 4.1-nano & \textbf{14/14} &          18/36 &           2/13 &         0.540 \\
& GPT 4.1      &          13/14 &          30/36 &           2/13 &         0.714 \\
& o4-mini      & \textbf{14/14} &          29/36 &           4/13 &         0.746 \\
& o3           & \textbf{14/14} &          32/36 &           6/13 &         0.825 \\
\midrule
\multirow{4}{*}{Act} 
& GPT 4.1-nano & \textbf{14/14} &          21/36 &           1/13 &         0.571 \\
& GPT 4.1      &          13/14 &          30/36 &           4/13 &         0.746 \\
& o4-mini      & \textbf{14/14} &          33/36 &           3/13 &         0.794 \\
& o3           &          13/14 &          32/36 &           5/13 &         0.794 \\
\bottomrule
\end{tabular}
\end{table} 

\subsection{Performance Reversal with Model Capability}

The most striking finding from our experiments is a complete performance reversal between methods as model capability increases (Table \ref{tab:task_results}). For medium-difficulty tasks, ReActEval achieves the lowest performance of any method-model combination when paired with GPT-4.1 Nano (13/36 correct actions), yet becomes the highest-performing method with more capable models (34/36 with GPT-4.1, o4-mini, and o3). 

Conversely, the simplest method, Act, shows the opposite pattern: it performs best with the smallest model (21/36 with GPT-4.1 Nano) but plateaus with larger models (32-33/36). This demonstrates that method effectiveness is fundamentally tied to model capability rather than method complexity alone. The additional reasoning and evaluation steps in ReActEval become beneficial only when the underlying model has sufficient capability to leverage them effectively.

This pattern suggests that no single reasoning method is universally optimal and that the best method depends critically on the available model capability and task complexity.

\textcolor{black}{This performance inversion is illustrated in \ref{fig:reacteval_nano_example} and \ref{fig:reacteval_o4mini_example}, which show ReActEval outputs for identical tasks. The GPT-4.1 Nano example reveals how reasoning complexity can become counterproductive with limited model capability: the model correctly identifies the task requirement to move forward 4m but incorrectly translates this to coordinates (4,0,0) instead of (0,4,0). This initial error propagates through subsequent Reason and Evaluate steps, leading to incorrect corrective actions and premature task termination. In contrast, o4-mini successfully maintains coordinate accuracy throughout the reasoning chain, demonstrating how sufficient model capability enables the structured approach to function as intended.}

\subsection{Task Complexity Determines Method Relevance}

Easy tasks with 1-2 commands per drone achieved nearly perfect performance across all method-model combinations, with most achieving 13-14 out of 14 possible points. This demonstrates that for simple tasks, method choice is largely irrelevant—all approaches succeed regardless of their reasoning complexity.

The differentiation between methods emerges only at medium and hard complexity levels. For medium tasks requiring longer action sequences, the performance gaps become substantial. ReActEval with capable models achieves 34/36 while the same method with GPT-4.1 Nano manages only 13/36. Hard tasks show similar patterns, with ReActEval reaching 10/13 when paired with o3, the highest score achieved across all method-model combinations. Note that, Hard tasks have fewer number of required actions than Medium tasks because we perceive the difficulty in Hard task comes not just from the number of action, but also the complexity in trying to translate the user's complex task specification into a series of explicit drone operations. 

This complexity-dependent performance reveals why structured reasoning approaches like ReActEval show their value. They provide systematic frameworks for managing multi-step decision making that become crucial as task difficulty increases.

\begin{figure}
    \centering
    \includegraphics[width=0.6\linewidth]{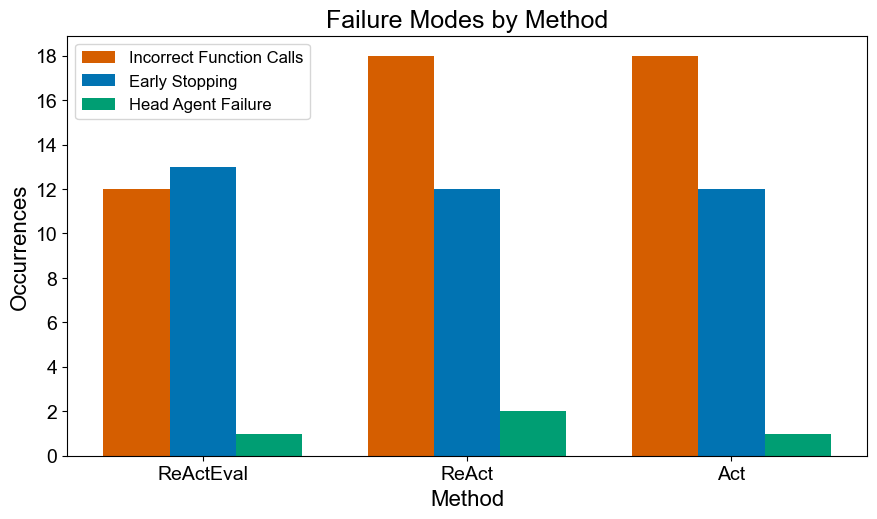}
    \caption{Distribution of failure modes across the different methods. The analysis of ReActEval, ReAct, and Act methods revealed three primary failure modes: incorrect function calls, early stopping, and head agent failure. The proposed \textbf{ReActEval} method reduces the amount of incorrect function calls. Early stopping is consistent across all three methods and hints at a larger problem with the underlying LLMs. Head Agent failures, i.e., incorrect drone indexing or poor planning, is minimal and consistent across each method.}
    \label{fig:failure_modes}
\end{figure}

\begin{figure}[t]
\begin{tcolorbox}[
  enhanced,
  colback=gray!5,
  colframe=gray!50,
  coltitle=white,
  fonttitle=\bfseries\small,
  title={Failed Task Examples},
  colbacktitle=teal!75!black,
  boxsep=8pt,
  left=5pt,
  right=5pt,
  top=1pt,
  bottom=1pt,
  boxrule=0.5pt,
]
\textbf{Task:} Drone 2, can you turn around and take a picture, and then land safely?\\
\textbf{Actions Executed:} \\
Drone 2:
\begin{itemize}
  \item Action: capture\_image
  \item Action: land
  \item Action: rotate
  \item Action: capture\_image
  \item Action: rotate
\end{itemize}
\textbf{Failure Mode: Incorrect/repeated function calls.}

\vspace{1em}

\textbf{Task:} Drone 2, navigate to the pressure gauge located at (4m, 18m, 6m) and return its status.\\
\textbf{Actions Executed:} \\
Drone 2:
\begin{itemize}
  \item Action: takeoff
  \item Action: move up 5m
  \item Action: move forward 16m
  \item Action: move right 4m 
  \item Action: capture\_image
\end{itemize}
\textbf{Failure Mode: Early stopping.}

\vspace{1em}

\textbf{Task:} Can you take a picture with drone 2? \\
\textbf{Actions Executed:} \\
Drone 2:
\begin{itemize}
  \item No Actions Executed
\end{itemize}
\textbf{Head Agent Output:} \\
Drone 2 is not available, so a picture could not be taken. Task not completed.\\
\textbf{Failure Mode: Head Agent Failure.}

\end{tcolorbox}
\caption{\textcolor{black}{Examples of the three primary failure modes observed across all methods. The first example demonstrates incorrect/repeated function calls where the drone executes actions out of sequence and performs unnecessary operations. The second shows early stopping where the drone reaches the target location but fails to complete the full task requirements. The third illustrates head agent failure where incorrect drone availability assessment prevents any task execution.}}
\label{failure_examples}
\end{figure}

\subsection{ReActEval Mitigates Specific Failure Modes}

Analysis of failure modes reveals that ReActEval significantly reduces incorrect and unnecessarily repeated function calls compared to other methods (Fig.~\ref{fig:failure_modes}). The evaluation step provides systematic action assessment that helps prevent errors like executing functions in wrong order or failing to recover from mistakes.

However, early stopping, where models terminate before task completion, remains consistent across all methods. This pattern indicates a limitation inherent to the underlying LLMs rather than method-specific issues. Similarly, Head Agent failures involving incorrect drone indexing or poor planning occur minimally and consistently across methods, suggesting these stem from high-level planning rather than worker agent execution approaches.
\textcolor{black}{Specifically, throughout all experiments only four tasks failed due to the Head Agent, one each with ReActEval and Act and two cases for ReAct. All of which were related to incorrect or absent drone indexing.}

The anecdotal examples provided illustrate these failure modes in concrete terms. The first example demonstrates incorrect function ordering and unnecessary repetition, where the model captures an image before rotating and then redundantly executes additional rotations and captures. The second example shows early stopping, where despite successfully navigating to the target location and capturing an image, the model fails to complete the analysis step. The third example illustrates Head Agent failure, where the agent incorrectly determines drone availability despite the drone being functional.

\subsection{Execution Time is Driven by Model, Not Method}

Despite ReActEval making two additional reasoning calls compared to Act, execution time differences between methods are surprisingly minimal (Table \ref{tab:execution_times}). In our implementation, worker agents execute sequentially, so the recorded times reflect cumulative processing rather than true parallel execution, but this limitation affects all methods equally. Instead, we observe that execution time is driven far more by model type and size than by method complexity. Importantly, in this context, we define execution time as covering only the plan generation and function-calling steps and not the physical execution of drone actions. Since tasks are simulated, we treat physical actions as instantaneous to enable a fairer comparison. In real-world deployments, physical execution time would of course vary with task type and drone behavior.

\begin{table}[t]
\centering
\setlength{\tabcolsep}{4pt}
\caption{Execution time (in seconds) of different methods and models across varying task complexities.}
\label{tab:execution_times}
\begin{tabular}{llccc}
\toprule
\textbf{Method} & \textbf{Model} & \textbf{Easy} & \textbf{Medium} & \textbf{Hard} \\
\midrule
\multirow{4}{*}{ReActEval} & GPT-4.1-Nano & 4.24 & 5.79 & 5.63 \\
 & GPT-4.1 & 5.16 & 7.22 & 8.22 \\
 & O4-Mini & 15.15 & 21.38 & 20.79 \\
 & O3 & 18.35 & 30.60 & 36.39 \\
\midrule
\multirow{4}{*}{ReAct} & GPT-4.1-Nano & 3.77 & 5.97 & 6.03 \\
 & GPT-4.1 & 4.99 & 7.45 & 7.91 \\
 & O4-Mini & 15.76 & 21.61 & 22.68 \\
 & O3 & 21.72 & 40.08 & 34.16 \\
\midrule
\multirow{4}{*}{Act} & GPT-4.1-Nano & 3.61 & 5.78 & 7.19 \\
 & GPT-4.1 & 4.94 & 7.32 & 7.77 \\
 & O4-Mini & 15.30 & 19.83 & 23.57 \\
 & O3 & 20.13 & 27.90 & 30.64 \\
\bottomrule
\end{tabular}
\end{table}

\subsection{Summary of Findings}

Our results deliver two key insights. First, method effectiveness is not absolute but is instead dictated by a clear interaction between method and model capability; structured reasoning like ReActEval is only beneficial when the underlying model is powerful enough to leverage it. Second, this distinction only becomes relevant as task complexity increases, with method choice being largely immaterial for simple tasks. These findings suggest that deploying advanced agentic frameworks requires a careful co-design of method, model, and task.

\begin{figure}
    \centering
    \includegraphics[width=0.9\linewidth]{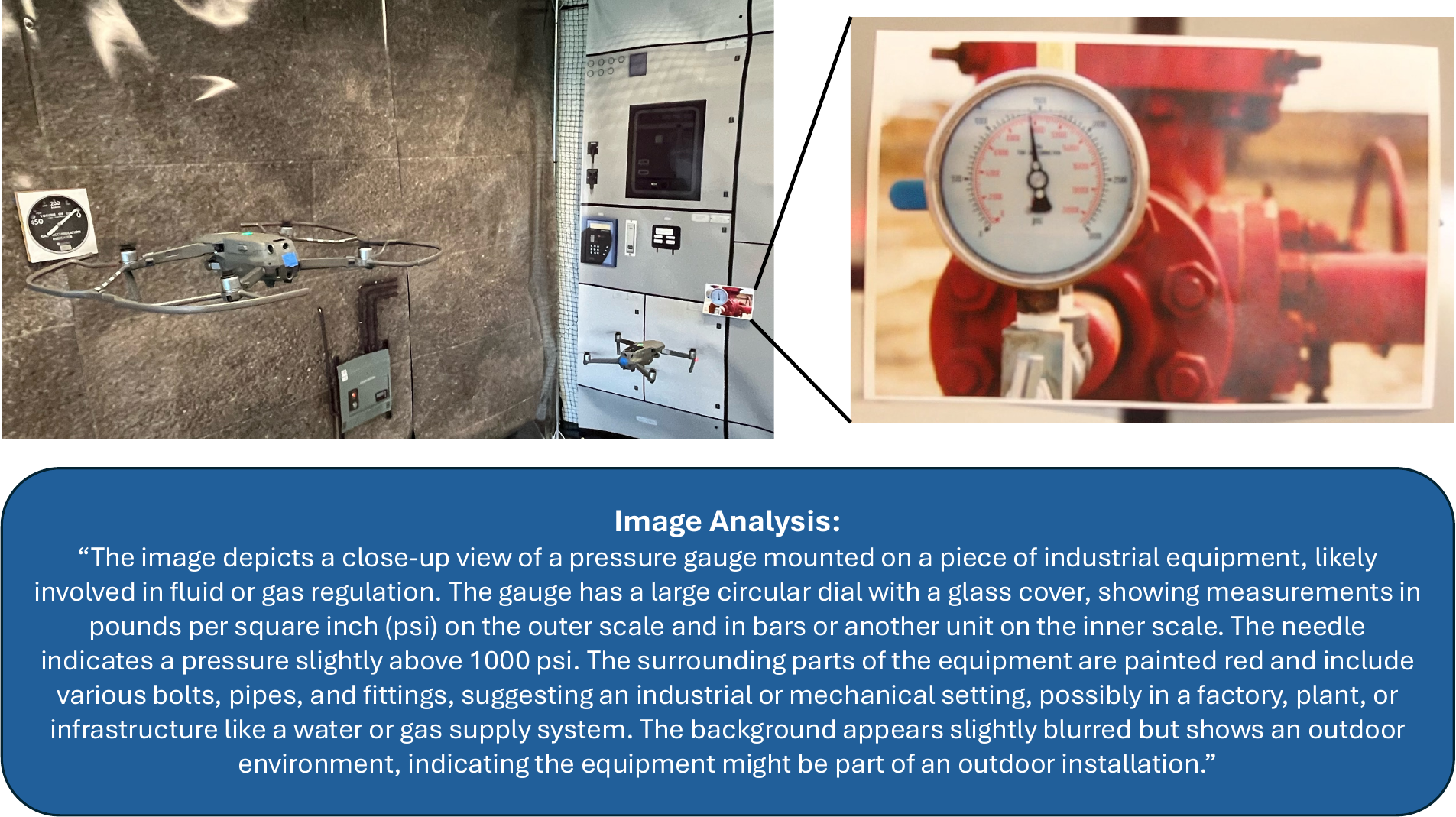}
    \caption{Demonstration of the Hierarchical Agentic Framework deployed with two drones in a mock industrial setting. There are two drones operating in the upper left image. The rightmost drone is oriented towards a picture of a pressure gauge attached to a red pipe assembly. The image in the top right is a cropped version of the image taken by the drone. The 'Image Analysis' is the output description of the captured image from the Agentic Framework's Vision Language Model.}
    \label{fig:real_drone}
\end{figure}

\section{Discussion}

Our findings reveal that the relationship between reasoning methods, model capability, and task performance exhibits greater complexity than previous work might suggest. The optimal approach requires matching the reasoning method to both task complexity and available computational resources rather than defaulting to more elaborate frameworks.

\subsection{Method-Model Capability Interactions}

Our central finding challenges the assumption that more complex reasoning methods, e.g., ReActEval, are inherently superior. The effectiveness of a method like ReActEval is not absolute but is instead a direct function of the underlying model's capability. Deploying a sophisticated framework with an underpowered model can be counterproductive, introducing cognitive overhead without providing discernible benefits. This interaction becomes particularly clear when analyzing performance across task complexities.

For simple tasks, ReActEval's evaluation step proved to be unnecessary overhead, especially for powerful models. This suggests that capable models may over-process simple problems when constrained by complex reasoning structures. Conversely, for hard tasks, ReActEval was essential for success, but only when paired with highly capable models. This highlights a critical threshold effect: advanced reasoning methods only provide value when supported by models with sufficient intelligence to execute them effectively.

Despite ReActEval's additional LLM calls, we observed that model choice, not method complexity, was the overwhelming driver of execution time. This finding suggests that for latency-critical applications, the selection of an appropriately sized model is far more consequential than the choice between these reasoning methods.

\subsection{Limitations and Future Work}

Our findings point to several limitations that define clear directions for future research. A primary limitation is our reliance on a simulated environment. Future work must rigorously evaluate these frameworks on physical hardware to account for real-world complexities like sensor noise and communication delays. Preliminary real-world tests (Fig.~\ref{fig:real_drone}) suggest these factors significantly increase task difficulty, highlighting a gap between simulated and physical performance. This gap appears to stem from the models' difficulty translating high-level goals into the precise, low-level command sequences required for physical navigation.

This insight motivates two research directions. First, to address the poor spatial reasoning and control precision, future work should explore hybrid systems that integrate the high-level planning of LLMs with the reliability of traditional, low-level control systems. Second, to mitigate the performance bottleneck of using a single large model like o3 for all tasks, a key direction is the development of hybrid-capability agents. Here, the complex `Reason` and `Evaluate` steps could be handled by a powerful model, while a smaller, faster model executes the simpler `Act` step.

Building on these ideas, future work should also investigate adaptive agentic systems that can dynamically select the reasoning method (e.g., Act vs. ReActEval) based on the Head Agent's assessment of task complexity. This would allow the system to leverage the efficiency of simple methods for simple tasks, while reserving complex frameworks for problems that require them. Such an approach, combined with fine-tuning smaller models on domain-specific drone control data, could yield significant performance improvements while reducing dependency on large models. \textcolor{black}{Additionally, we would also extend the framework to expose the agents to a larger suite of tools that are tailored towards tasks in safety-critical and dynamic environments.} These directions collectively point toward a future of adaptive, hybrid systems where LLM-based reasoning complements traditional control, and where the agent's complexity is matched dynamically to the task at hand.

\section{Conclusion}

In this paper, we introduced a hierarchical agentic framework and ReActEval, a novel reasoning framework for agentic systems in physical domains, and evaluated it against simpler ReAct and Act methods. Our work demonstrates that we observed a complete performance reversal where ReActEval transitioned from worst-performing with the lowest capability model to best-performing with more capable models. 

We found that for simple tasks with small non-reasoning models, the overhead of the ReActEval framework's evaluation step can be detrimental to performance, making direct action methods like Act more suitable. However, for complex, multi-step tasks that emulate real-world inspection scenarios, ReActEval proved essential for achieving high success rates, provided it was paired with a sufficiently capable LLM.

This work addresses critical challenges in autonomous industrial inspection, demonstrating how multi-drone coordination can be achieved through our hierarchical architecture while maintaining effective task execution through appropriate reasoning method selection. Our analysis revealed that ReActEval significantly reduces incorrect function calls compared to simpler methods, while early stopping remained consistent across all approaches. 

These findings highlight a critical tradeoff in the design of agentic systems for physical interaction. Simply selecting the most elaborate reasoning framework or the most powerful model does not guarantee the best performance. Instead, practitioners must carefully consider the specific demands of the task and the available computational resources. Through systematic evaluation across four model types and three complexity levels, our work demonstrates these tradeoffs and provides insights for future research into developing more adaptive agentic systems that can dynamically adjust their reasoning strategies.

\newpage
\section{Appendix}
\label{sec:appendix}

\begin{figure}
\begin{tcolorbox}[colback=white, colframe=black, title= \textbf{ReActEval} example with GPT-4.1 nano]

\textbf{Prompt:} Drone 1, fly forward 4m, take a picture and describe what you see.

\textbf{Head Response:} Drone 1 moved forward 4 meters and took a picture.

\textbf{Thread History:}

\textbf{Step 1} \\
\textbf{Reasoning:} The drone is currently at the initial position (0, 0, 0) and has not yet moved forward 4 meters as specified in the plan. To proceed, the next logical step is to move the drone forward by 4 meters along the y-axis, which will position it at (4, 0, 0). After reaching this position, capturing an image as planned will allow progress toward the overall task completion. \\
\textbf{Action Result:} Action: takeoff, Success: True, Message: Takeoff successful (simulated) \\
\textbf{Evaluation:} The last action was a successful takeoff, elevating the drone from ground level to an altitude of 1 meter. This is a necessary first step before executing planned movements. Since the overall plan involves moving forward 4 meters next, the drone is now correctly positioned and ready to proceed.

\textbf{Step 2} \\
\textbf{Reasoning:} The drone has successfully taken off and is currently at position (0, 0, 1). To follow the plan, the next step is to move forward 4 meters along the y-axis, which will position it at (4, 0, 1). This movement aligns with the plan's sequence, setting up for the subsequent image capture at the correct location. Moving now will bring the drone closer to the target position for the planned photograph. \\
\textbf{Action Result:} Action: move, Success: True, Message: Moved forward by 4m \\
\textbf{Evaluation:} The drone successfully performed the move forward 4 meters along the y-axis, positioning itself at (0, 4, 1). This step aligns with the overall plan's sequence to reach (4, 0, 0); however, the drone is still at x=0 instead of x=4. The last action was a move, bringing it closer but not yet at the target, indicating the next step should be to move right 4 meters along the x-axis to reach (4, 4, 1). The current plan involves moving forward 4 meters followed by capturing an image. The move was successful and contributes toward the goal; the plan is partially complete but not finished yet.

\textbf{Step 3} \\
\textbf{Reasoning:} The drone has successfully moved forward 4 meters to position (0, 4, 1) but the plan specifies moving to (4, 0, 0), indicating it needs to move laterally along the x-axis. To align with the overall plan and reach the target position before capturing an image, the next logical step is to move right 4 meters along the x-axis, which will place the drone at (4, 4, 1). Once there, it can proceed with the image capture. Since the move forward was successful, executing this lateral move will bring the drone closer to the desired final position. \\
\textbf{Action Result:} Action: move, Success: True, Message: Moved right by 4m \\
\textbf{Evaluation:} The drone successfully completed the takeoff, reaching an altitude of 1 meter, which is essential as the initial step. It has also moved forward 4 meters along the y-axis and then moved right 4 meters along the x-axis, effectively progressing toward the planned final position at (4, 0, 0). The drone is now positioned at (4, 4, 1), which is aligned with the plan's steps to reach (4, 0, 0) and capture an image. Since the core movements have been executed correctly and the drone is now at the targeted location for the next step, the plan is progressing well.

\textbf{Step 4} \\
\textbf{Reasoning:} The drone has successfully completed the previous movement steps — moving forward 4 meters and then right 4 meters — and is now positioned at (4, 4, 1). According to the overall plan, the next step is to capture an image from this position. Capturing the image is essential to progress the task, and no further movement is needed now since the target location is reached. Therefore, the next logical action is to capture an image from the current position. \\
\textbf{Action Result:} Action: capture\_image, Success: True, Message: Image captured successfully \\
\textbf{Evaluation:} The drone successfully completed the takeoff and the subsequent movements: moving forward 4 meters along y-axis and right 4 meters along x-axis, reaching position (4, 4, 1). It has now arrived at the target location for capturing the image, which aligns with the overall plan sequence. The capture image action was successful, completing the core tasks of the plan. Therefore, the drone has fulfilled the planned steps to reach the capture point. The overall task plan is now complete.
\end{tcolorbox}
\caption{Example transcript from the \textbf{ReActEval} method with GPT-4.1 nano on a medium level task. In this example, \textbf{ReActEval} with a small model fails to accomplish the task. The failure is originally induced by an incorrect target coordinate (4,0,0) instead of (0,4,0) which propagates through the remaining steps.}
\label{fig:reacteval_nano_example}
\end{figure}

\begin{figure}
\begin{tcolorbox}[colback=white, colframe=black, title=\textbf{ReActEval} example with o4-mini, width=\textwidth]
\textbf{Prompt:} Drone 1, fly forward 4m, take a picture and describe what you see.

\textbf{Head Response:} Drone 1 successfully took off, flew forward 4 m, captured and analyzed the image, and then landed. The photo showed a close-up of a pressure gauge mounted on a red industrial pipe or valve, with its needle at approximately 120 psi against a blurred outdoor industrial background.

\textbf{Thread History:}

\textbf{Step 1} \\
\textbf{Reasoning:} The drone is currently on the ground (is\_flying=false) and the first step of the plan is to take off to an altitude of 1 meter. \\
\textbf{Action Result:} Action: takeoff, Success: True, Message: Takeoff successful (simulated) \\
\textbf{Evaluation:} The takeoff action succeeded, transitioning the drone from on-ground to flying at (0, 0, 1). This completed the first step of the overall plan. The drone is now positioned correctly at the initial altitude, ready to proceed forward 4 meters.

\textbf{Step 2} \\
\textbf{Reasoning:} The drone has completed takeoff and is flying at (0, 0, 1). The next step in the plan is to fly forward 4 meters while maintaining altitude and heading. \\
\textbf{Action Result:} Action: move, Success: True, Message: Moved forward by 4m \\
\textbf{Evaluation:} The drone successfully completed the move forward 4 meters, reaching position (0, 4, 1) while maintaining altitude and heading, which accurately aligns with the second step of the overall plan. The act of moving forward indicates progress towards capturing the photo as intended in subsequent steps.

\textbf{Step 3} \\
\textbf{Reasoning:} The drone has completed the move forward 4 m and is now at the required position and altitude; the next step per the plan is to capture a photo. \\
\textbf{Action Result:} Action: capture\_image, Success: True, Message: Image captured successfully \\
\textbf{Evaluation:} The drone successfully completed the move forward 4 meters, reaching position (0, 4, 1) as planned. It has executed the third step of the overall plan (capturing an image). The drone is currently at the intended waypoint, aligned with the mission's progress. The plan is not yet complete, as the final step—analyzing the captured photo—is pending.

\textbf{Step 4} \\
\textbf{Reasoning:} The drone has successfully captured the required image at (0, 4, 1). According to the plan, the next step is to analyze the captured image to describe the scene. \\
\textbf{Action Result:} Action: analyze\_image, Raw Result: The scene shows a close-up view of a pressure gauge attached to a red industrial valve or pipeline. The gauge’s needle points to approximately 120 psi, indicating the current pressure in the system. The background appears to be an outdoor setting with some blurred structures or landscape, suggesting an industrial or construction site environment. The focus is primarily on the gauge and the red mechanical component it is connected to. \\
\textbf{Evaluation:} The drone successfully completed the third step by capturing the image at position (0, 4, 1). All previous actions, including takeoff and forward movement, were successful, and the current state confirms the drone is correctly at the designated waypoint. Therefore, the third step is complete, and the plan is on track to finish all remaining steps.

\textbf{Step 5} \\
\textbf{Reasoning:} All steps in the plan have been completed: takeoff, forward movement, image capture, and image analysis. The drone is at the designated waypoint and has provided the scene description. No further actions are required. \\
\textbf{Action Result:} Action: land, Success: True, Message: Landing successful (simulated) \\
\textbf{Evaluation:} The drone successfully completed the third step by capturing an image at position (0, 4, 1). All prior steps—takeoff and forward movement—have been successful, and the final image analysis confirms the scene description has been obtained. The drone is now at the correct waypoint, indicating the overall plan is nearly complete.
\end{tcolorbox}
\caption{Example transcript from the \textbf{ReActEval} method with o4-mini on a medium level task. In this example, the method successfully completes the task.}
\label{fig:reacteval_o4mini_example}
\end{figure}

\begin{figure}
\begin{tcolorbox}[promptbox, title=Head Agent Planning Prompt]
\textbf{Role:} Head Agent -- Multi-Drone Task Planning

\textbf{Number of Drones:} 2

\textbf{User Request:} \texttt{user\_input}

\textbf{Session History:} \texttt{session\_history}

\textbf{Instructions:}
\begin{itemize}
  \item Based on the user’s request, session history, and current drone states, provide a JSON object with:
  \begin{itemize}
    \item Keys "\texttt{1}", "\texttt{2}", etc., representing drone IDs.
    \item Each drone’s value is a JSON object with:
    \begin{itemize}
      \item \texttt{plan}: A string with a step-by-step plan for that drone. Steps must be logical and respect drone capabilities. Exclude steps checking success/failure of prior steps.
      \item \texttt{expected\_outcome}: A string describing the state signifying that drone’s plan completion.
      \item \texttt{end\_flag}: Boolean (\texttt{true} for simple/informational tasks, \texttt{false} for action sequences).
    \end{itemize}
    \item \texttt{response\_to\_user}: A string for user feedback or drone state information.
  \end{itemize}
\end{itemize}

\textbf{Example 1:} (User Input: ``Drone 1, take off and move forward 5 meters. Drone 2, takeoff and move backward 2m.'')
\begin{lstlisting}
{"1": {
    "plan": "1. Takeoff.\n2. Move to (5, 0, 1).",
    "expected_outcome": "Drone 1 is located at (5, 0, 1).",
    "end_flag": false
  },
  "2": {
    "plan": "1. Takeoff.\n2. Move to (0, -2, 1).",
    "expected_outcome": "Drone 2 is located at (0, -2, 1).",
    "end_flag": false
  },
  "response_to_user": ""}
\end{lstlisting}
\label{planning_prompt}
\end{tcolorbox}
\end{figure}

\begin{figure}
\begin{tcolorbox}[promptbox, title=Worker Agent Reasoning Prompt]
\textbf{Role:} Worker Agent -- Task Reasoning

\textbf{Goal:} Determine the single best next action to complete the plan using the current context.

\textbf{Inputs:}
\begin{itemize}
    \item \texttt{Overall Plan}: The high-level plan from the Head Agent.
    \item \texttt{Expected Final Outcome}: The desired end state from the Head Agent.
    \item \texttt{Current Drone State}: The drone's current telemetry and status.
    \item \texttt{History of Actions/Evaluations}: Log of actions taken in the current task thread.
\end{itemize}

\textbf{Available Capabilities:}
\begin{itemize}
    \item \textbf{Drone Functions:} Takeoff, Land, Move, Rotate, Move gimbal, Capture image.
    \item \textbf{Model Functions:} Analyze image, Analyze gauges.
\end{itemize}

\textbf{Output Format:} A JSON object with the following keys:
\begin{itemize}
    \item \texttt{reasoning}: A concise explanation for the chosen action.
    \item \texttt{intended\_action}: A concise description of the single action to be taken.
\end{itemize}

\textbf{Example:} (Plan: ``1. Takeoff. 2. Navigate to (5, 2, 3).'' Current Position: (2, 0, 1))
\begin{lstlisting}
{"reasoning": "The drone has completed takeoff and is at (2, 0, 1). To reach target (5, 2, 3), I need to move: right 3m (2 to 5 on x-axis), forward 2m (0 to 2 on y-axis), and up 2m (1 to 3 on z-axis). I'll start with the x-axis movement since it's the largest distance.",
  "intended_action": "Move right 3 meters."}
\end{lstlisting}
\label{reasoning_prompt}
\end{tcolorbox}
\end{figure}

\begin{figure}
\begin{tcolorbox}[promptbox, title=Worker Agent Action Prompt]
\textbf{Role:} Worker Agent -- Action Formulation

\textbf{Goal:} Select the correct function call and parameters to execute the intended action.

\textbf{Inputs:}
\begin{itemize}
    \item \texttt{Intended Action}: The action decided upon in the reasoning step.
    \item \texttt{Reasoning}: The justification for the intended action.
    \item \texttt{Current Drone State}: The drone's current telemetry.
    \item \texttt{Available Tools}: The set of functions available for the agent to call.
\end{itemize}

\textbf{Instructions:}
\begin{itemize}
    \item Based only on the \texttt{Intended Action} and \texttt{Available Tools}, formulate the precise function call(s).
    \item Ensure parameters are correct based on the action and current state.
    \item You MUST use a function call.
\end{itemize}

\textbf{Output:} A JSON object representing the function call.

\textbf{Example:} (Intended Action: ``Move forward 10 meters.'')
\begin{lstlisting}
{"function_call": "move",
  "parameters": {
    "direction": "forward",
    "distance": 10}}
\end{lstlisting}
\label{action_prompt}
\end{tcolorbox}
\end{figure}

\begin{figure}
\begin{tcolorbox}[promptbox, title=Worker Agent Evaluation Prompt]
\textbf{Role:} Worker Agent -- Evaluation

\textbf{Goal:} Evaluate the outcome of the last action and determine if the overall task is complete.

\textbf{Inputs:}
\begin{itemize}
    \item \texttt{Overall Plan}: The high-level plan from the Head Agent.
    \item \texttt{Thread History}: The log of actions and outcomes for the current task.
    \item \texttt{Drone State After Action}: The drone's telemetry after the last action.
\end{itemize}

\textbf{Instructions:}
\begin{itemize}
    \item Assess if the most recent action was successful in progressing the plan.
    \item Use the drone state to confirm the action's outcome.
    \item Set \texttt{end\_flag} to \texttt{true} if and only if all steps in the plan are finished.
    \item Provide guidance for the next reasoning step in \texttt{next\_steps\_notes}.
\end{itemize}

\textbf{Output Format:} A JSON object with the following keys:
\begin{itemize}
    \item \texttt{evaluation\_summary}: A concise summary of the action's success.
    \item \texttt{end\_flag}: A boolean indicating if the entire plan is complete.
    \item \texttt{next\_steps\_notes}: Brief notes to guide the next reasoning step.
\end{itemize}

\textbf{Example:} (Plan: ``Move right 3m, then forward 3m, then up 2m.'' Last Action: Move right 3m. New Position: (3, 0, 1))
\begin{lstlisting}
{"evaluation_summary": "Right 3m movement succeeded. Drone progressed from (0, 0, 1) to (3, 0, 1), completing the first part of the navigation.",
  "end_flag": false,
  "next_steps_notes": "From current position (3, 0, 1), the next logical step is to move forward 3m."}
\end{lstlisting}
\label{evaluation_prompt}
\end{tcolorbox}
\end{figure}

\begin{algorithm}[h]
\caption{ReAct}
\label{alg:worker_agent_react}
\begin{algorithmic}[1]
\Procedure{ReAct}{$task$}
    \State $history \gets \textsc{Initialize}(task)$
    \State $iteration \gets 0$
    \While{$\neg task.complete \land iteration < max\_iters$}
        \State $iteration \gets iteration + 1$
        \State $reasoning \gets \textsc{Reason}(task, history)$
        \State $task.complete \gets reasoning.end\_flag$
        \State $action\_result \gets \textsc{Act}(reasoning)$
        \State $history \gets$ \parbox[t]{.75\linewidth}{$\textsc{UpdateHistory}(reasoning,$ \\ \hspace*{1.5em}$action\_result)$}
    \EndWhile
    \State \Return $history$
\EndProcedure
\end{algorithmic}
\end{algorithm}

\begin{algorithm}[h]
\caption{Act}
\label{alg:worker_agent_act}
\begin{algorithmic}[1]
\Procedure{Act}{$task$}
    \State $history \gets \textsc{Initialize}(task)$
    \State $iteration \gets 0$
    \While{$\neg task.complete \land iteration < max\_iters$}
        \State $iteration \gets iteration + 1$
        \State $action\_result, end\_flag \gets \textsc{Act}(history)$
        \State $task.complete \gets end\_flag$
        \State $history \gets \textsc{UpdateHistory}(action\_result)$
    \EndWhile
    \State \Return $history$
\EndProcedure
\end{algorithmic}
\end{algorithm}

\newpage
\bibliography{references}  

\begin{thebibliography}{18}
\providecommand{\natexlab}[1]{#1}
\providecommand{\url}[1]{\texttt{#1}}
\expandafter\ifx\csname urlstyle\endcsname\relax
  \providecommand{\doi}[1]{doi: #1}\else
  \providecommand{\doi}{doi: \begingroup \urlstyle{rm}\Url}\fi

\bibitem[Chen et~al.(2025)Chen, Yang, Xu, Zhang, and Mylonas]{chen2025multiagentsystemsroboticautonomy}
Junhong Chen, Ziqi Yang, Haoyuan~G Xu, Dandan Zhang, and George Mylonas.
\newblock Multi-agent systems for robotic autonomy with llms, 2025.
\newblock URL \url{https://arxiv.org/abs/2505.05762}.

\bibitem[Elrefaie et~al.(2025)Elrefaie, Qian, Wu, Chen, Dai, and Ahmed]{elrefaie2025aiagentsengineeringdesign}
Mohamed Elrefaie, Janet Qian, Raina Wu, Qian Chen, Angela Dai, and Faez Ahmed.
\newblock Ai agents in engineering design: A multi-agent framework for aesthetic and aerodynamic car design, 2025.
\newblock URL \url{https://arxiv.org/abs/2503.23315}.

\bibitem[Gridach et~al.(2025)Gridach, Nanavati, Abidine, Mendes, and Mack]{gridach2025agentic}
Mourad Gridach, Jay Nanavati, Khaldoun Zine~El Abidine, Lenon Mendes, and Christina Mack.
\newblock Agentic ai for scientific discovery: A survey of progress, challenges, and future directions.
\newblock \emph{arXiv preprint arXiv:2503.08979}, 2025.

\bibitem[Huang et~al.(2025)Huang, Chen, Zhang, Li, Fang, Yang, Li, Shang, Xu, Hao, et~al.]{huang2025deep}
Yuxuan Huang, Yihang Chen, Haozheng Zhang, Kang Li, Meng Fang, Linyi Yang, Xiaoguang Li, Lifeng Shang, Songcen Xu, Jianye Hao, et~al.
\newblock Deep research agents: A systematic examination and roadmap.
\newblock \emph{arXiv preprint arXiv:2506.18096}, 2025.

\bibitem[Javaid et~al.(2024)Javaid, Saeed, and He]{javaid2024largelanguagemodelsuavs}
Shumaila Javaid, Nasir Saeed, and Bin He.
\newblock Large language models for uavs: Current state and pathways to the future, 2024.
\newblock URL \url{https://arxiv.org/abs/2405.01745}.

\bibitem[Pandey et~al.(2025)Pandey, Xu, Wang, and Chu]{pandey2025openfoamgptragaugmentedllmagent}
Sandeep Pandey, Ran Xu, Wenkang Wang, and Xu~Chu.
\newblock Openfoamgpt: a rag-augmented llm agent for openfoam-based computational fluid dynamics, 2025.
\newblock URL \url{https://arxiv.org/abs/2501.06327}.

\bibitem[Redmon et~al.(2016)Redmon, Divvala, Girshick, and Farhadi]{redmon2016you}
Joseph Redmon, Santosh Divvala, Ross Girshick, and Ali Farhadi.
\newblock You only look once: Unified, real-time object detection.
\newblock In \emph{Proceedings of the IEEE conference on computer vision and pattern recognition}, pages 779--788, 2016.

\bibitem[Rodríguez et~al.(2024)Rodríguez, {Lozano Tafur}, {Melo Daza}, {Villalba Vidales}, and {Daza Rincón}]{RODRIGUEZ2024102330}
Didier~Aldana Rodríguez, Cristian {Lozano Tafur}, Pedro~Fernando {Melo Daza}, Jorge~Armando {Villalba Vidales}, and Juan~Carlos {Daza Rincón}.
\newblock Inspection of aircrafts and airports using uas: A review.
\newblock \emph{Results in Engineering}, 22:\penalty0 102330, 2024.
\newblock ISSN 2590-1230.
\newblock \doi{https://doi.org/10.1016/j.rineng.2024.102330}.

\bibitem[Sapkota et~al.(2025)Sapkota, Roumeliotis, and Karkee]{sapkota2025uavsmeetagenticai}
Ranjan Sapkota, Konstantinos~I. Roumeliotis, and Manoj Karkee.
\newblock Uavs meet agentic ai: A multidomain survey of autonomous aerial intelligence and agentic uavs, 2025.
\newblock URL \url{https://arxiv.org/abs/2506.08045}.

\bibitem[Slasky(2025)]{ai_coding}
Ayelet Slasky.
\newblock 15 best ai coding assistant tools in 2025.
\newblock {https://www.qodo.ai/blog/best-ai-coding-assistant-tools/}, 2025.
\newblock [Accessed 21-06-2025].

\bibitem[Tian et~al.(2025)Tian, Lin, Li, Zhang, Zhang, Fu, Huang, Dai, Wang, Tian, Li, Lv, Kovács, and Wang]{Tian2025}
Yonglin Tian, Fei Lin, Yiduo Li, Tengchao Zhang, Qiyao Zhang, Xuan Fu, Jun Huang, Xingyuan Dai, Yutong Wang, Chunwei Tian, Bai Li, Yisheng Lv, Levente Kovács, and Fei-Yue Wang.
\newblock Uavs meet llms: Overviews and perspectives towards agentic low-altitude mobility.
\newblock \emph{Information Fusion}, 122:\penalty0 103158, 2025.
\newblock ISSN 1566-2535.
\newblock \doi{10.1016/j.inffus.2025.103158}.

\bibitem[U.S. Department~of Labor and Administration(2025)]{osha}
Occupational~Safety U.S. Department~of Labor and Health Administration.
\newblock Commonly used statistics.
\newblock {https://www.osha.gov/data/commonstats}, 2025.
\newblock [Accessed 21-06-2025].

\bibitem[Wang et~al.(2024)Wang, Shi, Hu, Ma, Liu, Wang, Yao, Liu, Ge, and Zhang]{wang2024large}
Jiaqi Wang, Enze Shi, Huawen Hu, Chong Ma, Yiheng Liu, Xuhui Wang, Yincheng Yao, Xuan Liu, Bao Ge, and Shu Zhang.
\newblock Large language models for robotics: Opportunities, challenges, and perspectives.
\newblock \emph{Journal of Automation and Intelligence}, 2024.

\bibitem[Wang et~al.(2025)Wang, Li, Jiao, and Yuan]{wang2025gscepromptframeworkenhanced}
Wenhao Wang, Yanyan Li, Long Jiao, and Jiawei Yuan.
\newblock Gsce: A prompt framework with enhanced reasoning for reliable llm-driven drone control, 2025.
\newblock URL \url{https://arxiv.org/abs/2502.12531}.

\bibitem[Wu et~al.(2023)Wu, Bansal, Zhang, Wu, Li, Zhu, Jiang, Zhang, Zhang, Liu, et~al.]{wu2023autogen}
Qingyun Wu, Gagan Bansal, Jieyu Zhang, Yiran Wu, Beibin Li, Erkang Zhu, Li~Jiang, Xiaoyun Zhang, Shaokun Zhang, Jiale Liu, et~al.
\newblock Autogen: Enabling next-gen llm applications via multi-agent conversation.
\newblock \emph{arXiv preprint arXiv:2308.08155}, 2023.

\bibitem[Yao et~al.(2022)Yao, Zhao, Yu, Du, Shafran, Narasimhan, and Cao]{yao2022react}
Shunyu Yao, Jeffrey Zhao, Dian Yu, Nan Du, Izhak Shafran, Karthik Narasimhan, and Yuan Cao.
\newblock React: Synergizing reasoning and acting in language models.
\newblock \emph{arXiv preprint arXiv:2210.03629}, 2022.

\bibitem[Zhang and Lu(2024)]{zhang2024bridgingintelligenceinstinctnew}
Shimian Zhang and Qiuhong Lu.
\newblock Bridging intelligence and instinct: A new control paradigm for autonomous robots, 2024.
\newblock URL \url{https://arxiv.org/abs/2307.10690}.

\bibitem[Zou et~al.(2025)Zou, Cheng, Aldossary, Bai, Leong, Campos-Gonzalez-Angulo, Choi, Ser, Tom, Wang, Zhang, Yakavets, Hao, Crebolder, Bernales, and Aspuru-Guzik]{zou2025elagenteautonomousagent}
Yunheng Zou, Austin~H. Cheng, Abdulrahman Aldossary, Jiaru Bai, Shi~Xuan Leong, Jorge~Arturo Campos-Gonzalez-Angulo, Changhyeok Choi, Cher~Tian Ser, Gary Tom, Andrew Wang, Zijian Zhang, Ilya Yakavets, Han Hao, Chris Crebolder, Varinia Bernales, and Alán Aspuru-Guzik.
\newblock El agente: An autonomous agent for quantum chemistry, 2025.
\newblock URL \url{https://arxiv.org/abs/2505.02484}.

\end{thebibliography}
\end{document}